\documentclass[aps,prc,preprint,groupedaddress,showpacs]{revtex4}

\usepackage{graphicx}

\begin{document}

\title{Effect of neutron composition and excitation energy of primary fragments on isospin observables in 
          multifragmentation}
\author{D.V. Shetty, A.S. Botvina\footnote{On leave from Institute for Nuclear Research, 
               117312 Moscow, Russia.}, S.J. Yennello, A. Keksis, E. Martin, and G.A. Souliotis}
\affiliation{Cyclotron Institute, Texas A\&M University, College Station, Texas 77843, USA}
\date{\today}

\begin{abstract}
The isospin properties of primary and secondary fragments produced in 
multifragmentation of  Fe + Ni and Fe + Fe systems with respect to Ni + Ni system 
are analyzed within the statistical multifragmentation model framework.
The reduced neutron and proton densities show an asymmetry in the primary 
fragments, that is lessened after secondary decay. With increasing isospin (N/Z) 
this effect increases, while the sensitivity of fragment isospin toward excitation energy 
and N/Z of the primary fragments remains unchanged. 
\end{abstract}


\maketitle

\section{Introduction}
Recently, the importance of the isospin degree of freedom in obtaining 
information regarding charge equilibration and the symmetry dependent 
terms of the nuclear equation of state has prompted measurements of 
the isotopic distributions of emitted particles in multifragmentation process
\cite{XU00,MIL00,TSA01,TSAN01,TSA02,YEN94,JOH96,JOH97,ROW03}. 
These particles, when emitted, have a broad range of isospin content and 
carry important information regarding the evolution of the isospin (N/Z) degree 
of freedom. In multifragmentation, the system quickly expands 
into the vacuum and decays into hot primary  fragments. These hot fragments, despite their short 
life, carry a significant amount of information about the early stages of the reaction. 
The surface properties of the primary fragments may provide information about the 
critical temperature of nuclear matter  \cite{OGU02, KAR03}, whereas their 
isospin is related to the symmetry term in the fragment 
(nuclear liquid) phase \cite{BOT02, ONO03}. Similar hot nuclei are also produced in physical processes, 
such as the collapse of massive stars following supernova II explosions \cite{BETHE}, 
where they can live for a long time in equilibrium with the 
surrounding nuclear matter.
\par
Experimentally, the observed fragments result from the secondary 
decay of the hot primary fragments. 
The secondary decay smears out some of the relevant information 
making it difficult to study any isospin effect that would have been 
present before the secondary decay. It is important to 
verify the extent to which isospin information is preserved in 
the experimentally measured fragment observables. 
Therefore, a  self-consistent theoretical calculation of primary and secondary 
fragments, which includes a comparison to many  experimental 
observables, is necessary. Also, the secondary de-excitation process  
proceeds in physical conditions different from the standard de-excitation 
of a compound nucleus at low excitation energies: the primary fragments 
 are in the Coulomb field of other fragments. 
The structure of their excited levels can thus be different. 
These difficulties regarding the secondary de-excitation can be overcome  
by comparing the same 
observables in two similar reactions that differ mainly in isospin 
asymmetry (see e.g.\cite{XU00}). Such an observable can also be 
constructed from the model calculation and compared with experimental 
results. In a previous study \cite{SHE03}, assuming statistical 
equilibrium in an excited nuclear system, reduced ('free') 
neutron ($\rho_n$) and proton ($\rho_p$) densities were obtained from the 
measured isotope and isotone yield ratios. It was shown 
that  $\rho_n$ and $\rho_p$ depend significantly on the beam energy, 
and therefore, may carry  information about evolution of 
fragment isospin with excitation energy. In this work,
we show that the secondary decay decreases the asymmetry in the 
reduced neutron and proton densities and that this effect is larger 
for  system with higher isospin. The decay, however,  still 
preserves the sensitivity of the neutron content of fragments toward the excitation 
energy and N/Z of the total system. The observed asymmetry 
can be well accounted for by the statistical multifragmentation model 
(SMM) \cite{BON95}. 
\section{Experiment} 
The measurements were performed at the Cyclotron Institute of Texas A$\&$M 
University (TAMU) using beams provided by the K500
Superconducting Cyclotron. Isotopically pure beams of $^{58}$Ni and 
$^{58}$Fe at 30, 40 and 47 MeV/nucleon were bombarded on self-supporting 
$^{58}$Ni (1.75 mg/cm$^{2}$) and $^{58}$Fe (2.3 mg/cm$^{2}$) targets. Six 
discrete particle telescopes placed inside a scattering chamber 
and centered at laboratory angles of 10$^{\circ}$, 44$^{\circ}$, 
72$^{\circ}$, 100$^{\circ}$, 128$^{\circ}$ and 148$^{\circ}$, were used 
to measure fragments from the reactions. Each telescope consisted of a gas 
ionization chamber (IC) followed by a pair of silicon detectors 
(Si-Si) and a CsI scintillator detector, providing three distinct detector 
pairs (IC-Si, Si-Si, and Si-CsI) for fragment identification. Further details of the 
experimental setup and analysis can be found in refs. \cite{SHE03,RAM98}.
The present study is carried out for fragments detected at 44 degrees in 
the laboratory system, which corresponds to the center of mass angle 
$\approx 90$ degrees. The fragments detected at this angle originate 
predominantly from central events. Analysis with QMD+SMM 
model carried out in Ref. \cite{RAM98} supports this conclusion. 
Nevertheless, a possible admixture of the projectile/target 
fragmentation processes will be accounted in the following analysis. 
\section{Ratios of reduced nucleon densities} 
In the Grand-Canonical approach for the multifragmentation process 
(see e.g. \cite{BON95,RAN81,ALB85,BOT87}), the yield for an isotope with neutron 
number, N, and proton number, Z (the mass number A = N + Z), can be written as 

\begin{equation}
 Y(N,Z) \propto V \rho_{n}^{N}\rho_{p}^{Z}Z_{N,Z}(T)A^{3/2}e^{B(N,Z)/T}
\end{equation}

where V is the volume of the system,  $\rho_{n}$ $\propto$ e$^{\mu_{n}/T}$ 
and $\rho_{p}$ 
$\propto$ e$^{\mu_{p}/T}$ are the primary 'free' neutron and 
proton densities, respectively. 
$\mu_{n}$ and $\mu_{p}$ are the neutron and 
proton chemical potentials. $Z_{N,Z}(T)$ is the intrinsic partition 
function of the excited fragment. 
$B(N,Z)$ is the ground state binding energy of the corresponding fragment, 
and $T$ is the temperature. In the above formula, the effect of 
the Coulomb interaction on fragment yields is disregarded. 
In addition, nuclear systems are finite, and this fact influences 
the fragment yields, especially at low excitation energies \cite{BON95}. 
By introducing $\rho_{n}$ and $\rho_{p}$ the actual isotope yields are 
reduced to the approximation appropriate for the thermodynamical limit at 
high excitation energies \cite{BON95}, and hence referred to as the  'reduced' 
densities. 
\par
By taking ratios of isotope yields of two different 
systems, one can essentially decrease the uncertainties in the individual 
isotope yields which exist both in experimental measurements and 
their theoretical descriptions. In this way, the study of the isotopic composition 
of the fragments can be reduced to determine the relative neutron and proton densities 
obtained from the ratios of measured isotopes. The expression for the relative neutron density is 
\begin{equation}  
   \frac{ Y(N + k, Z)/Y^{Ni}(N + k, Z)} { Y(N, Z)/Y^{Ni}(N, Z)  } = 
\bigglb( \frac{\rho_n}{\rho_{n}^{Ni}} \biggrb )^{N} ,
\end{equation}
where $k$ corresponds to the different isotopes used to determine the double 
ratio, and $Y^{Ni}$ is the yield for the Ni + Ni reaction. A similar expression for 
the relative proton density can also be defined. Fig. 1 shows 
the average ratios of reduced neutron and proton densities as a 
function of the beam energy for Fe + Ni and Fe + Fe systems. 
The different symbols in the figure correspond to densities obtained for 
$k$ = 1, $k$ = 2 and $k$ = 3. One observes a steady decrease in the 
neutron density and a rise in the proton density with increasing beam 
energy for the Fe + Ni system. The effect is stronger for Fe + Fe 
system, having higher N/Z.
\section{Evaluation of parameters of the thermal sources}
In order to compare the results of Fig.1 with the statistical 
multifragmentation model, the initial parameters of the thermal source, such as 
the mass (A), charge (Z) and excitation energy ($E^{*}$), were evaluated by 
performing a dynamical BNV calculation \cite{BAR02}. 
For example, for Fe + Ni central collisions the parameters obtained for the 
three beam energies were $E^{*}_s\approx$ 5, 7 and 9.4 MeV/nucleon, 
$A_s\approx$ 111 , $Z_s\approx$ 52. In all cases the N/Z ratios of the 
sources are approximately the same as in interacting nuclei. These results 
were obtained 
    \begin{figure}
    \includegraphics[width=0.5\textwidth,height=0.50\textheight]{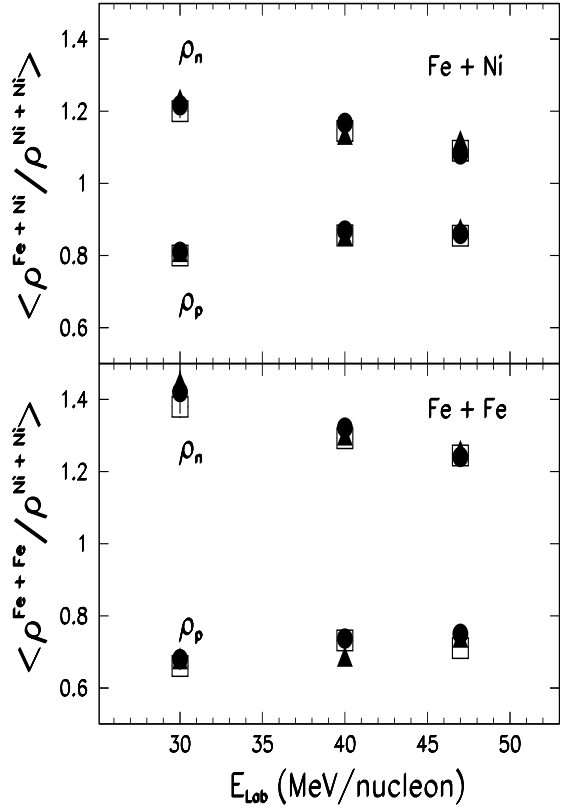} 
    \caption{Relative reduced neutron and proton densities as a function of 
                 bombarding energy for Fe + Ni and Fe + Fe reactions. The errors bars are of 
                 the size of symbols. The different symbols are explained in the text.}
   \end{figure}
at a time around 40-50 fm/c after  the projectile fuses with the target 
nuclei, and the quadrupole moment of the nucleon coordinates 
(used for identification of the deformation of the system) approaches zero. 
We mention that in the calculations, the form of single sources 
oscillates and they do not undergo dynamical disintegration into 
fragments. Consequently, there is no dynamical flow produced in 
these reactions. 
\par
Alternatively, the thermal source parameters can also be estimated from other 
experimental data. For example, in Ref.~\cite{HUD03}, for nearly symmetric 
central collisions of Xe + Sn, excitation energies of 
$E^{*}\approx$ 5 MeV/nucleon at 32 MeV/nucleon, and 
$E^{*}\approx$ 7 MeV/nucleon at 50 MeV/nucleon were obtained. 
It was shown in this work that the SMM describes the disassembly 
of these nuclear systems into fragments quite well, and 
reproduces the excitation energies of the observed primary fragments. 
However, in this case of heavy systems a radial flow 
of $\approx$ 2 MeV/nucleon was identified at 50 MeV/nucleon beam energy. 
In the present case, the systems are nearly twice as small and more 
stable with respect to excitation and flow. As supported by 
experimental observations (see e.g. \cite{HON98}), small systems produced 
in central collisions accumulate smaller radial flow than the large systems, 
consequently, a larger part of their available energy can be transformed into 
thermal excitation. 
\section{SMM calculation} 
The SMM calculation was performed for different sources in order to cover 
the uncertainties of the thermal source 
parameters. In particular, we selected `complete fusion' sources: $A_s$ = 116, 
with $Z_s$ = 52, 54, and 56, at excitation energies of 
    \begin{figure}
    \includegraphics[width=0.5\textwidth,height=0.50\textheight]{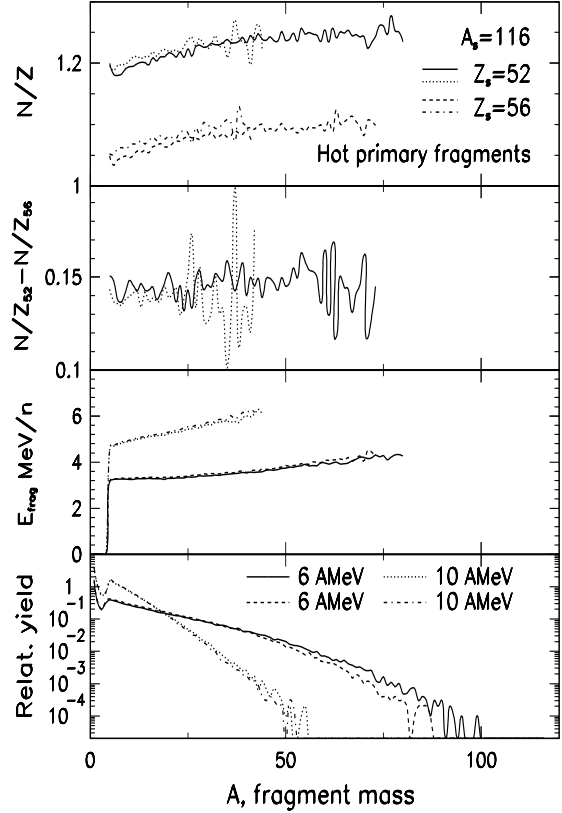} 
    \caption{Characteristics of hot primary fragments versus their 
                 mass number A, produced in the freeze-out volume during multifragmentation 
                 of the thermal sources with mass number $A_s$ = 116, and charges $Z_s$ = 52 and 
                 $Z_s$ = 56, at excitation energies $E^{*}_s$ = 6 and 10 MeV/nucleon (see 
                  notations in the figure). Panels from top to bottom: neutron to proton (N/Z) ratio; 
                 difference of N/Z ratios between neutron rich ($Z_s$ = 52) and neutron poor ($Z_s$ = 56) 
                 sources at different $E^{*}_s$; internal excitation energies; relative mass yields.} 
     \end{figure}
     \begin{figure}
     \includegraphics[width=0.5\textwidth,height=0.50\textheight]{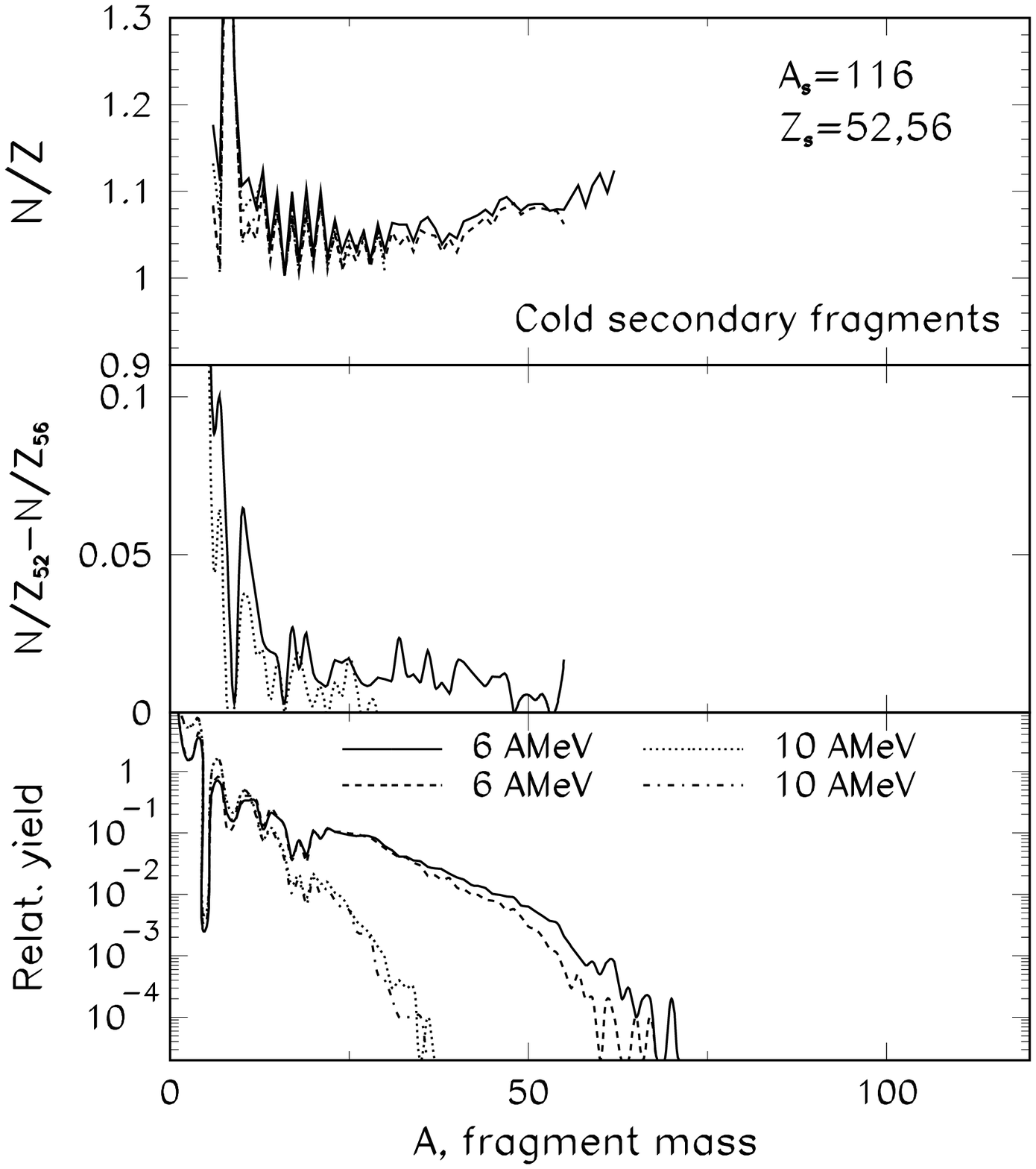}
     \caption{The same as in Fig.~2, but without the internal excitation 
                 panel, for the cold fragments produced after the secondary de-excitation.}
     \end{figure}
$E^{*}_s$ = 6, 8, and 10 MeV/nucleon. In order to  account for 
a possible loss of nucleons during pre-equilibrium emission, and 
investigate the sensitivity of the results toward the source size, 
we have also considered the sources $A_s$ = 58, with $Z_s$ = 26, 27, and 28. 
The small sources can also simulate the effects of a two source admixture 
that leads to the observable isospin characteristics. 
We have used the standard parameters for the SMM calculations, where the 
freeze-out density is taken as 1/3 of the normal nuclear density. 
This parametrization, together with the adopted secondary de-excitation 
procedure \cite{BOT87}, provide a  reasonably good description of the experimental data, 
including the isotope production (see e.g. \cite{MIL00,BOT95,WIN01}). 
We believe that our analysis of the reduced neutron and proton densities 
is not very sensitive to the previously mentioned difficulties related to the
treatment of secondary de-excitation processes: 
Since these densities are obtained by averaging over a large number of isotopes, 
the results are not affected by the rare isotopes which are produced 
with small probabilities. 
\par
In Fig. 2 we show the mean characteristics of the hot primary fragments 
produced after disintegration of the thermal sources in the freeze-out 
volume. We show their relative yields, internal excitation energies and 
N/Z ratios. Even at the lowest $E^{*}_s$ the systems undergo intensive multifragmentation, 
characterized by an exponential-like mass distribution. As expected, with 
increasing excitation the sources break-up into smaller 
fragments. At $E^{*}_s$ = 6 AMeV the internal excitations of the fragments are 
around 3 MeV/nucleon, and they increase slightly with their masses. 
This trend exists because the microcanonical calculation conserves 
energy: the channels with production of large fragments have smaller 
Q-values, and, consequently, higher temperatures. In case of canonical 
calculations, when the temperature is the same for all channels, the internal 
excitation energies per nucleon would be nearly the same for all fragments 
having the same level density structure. With increasing $E^{*}_s$ the 
internal excitation of fragments increases by $\sim$1.5 MeV/nucleon. 
By comparing the sources with different neutron content, we see that 
the main difference is in the N/Z ratios of primary fragments. However, 
these N/Z ratios change only slightly with the excitation energies. 
They are mainly determined by the N/Z of the sources (actually they 
are only slightly less than the sources' ratios), and they demonstrate 
a well-known statistical trend: The N/Z of fragments increases with their 
mass number \cite{BOT01}, which can be explained by the Bethe-Weizsaecker 
formula. Most of the neutrons of the system are accumulated in the 
hot fragments, the share of free primary neutrons is rather small. 
However, if we take the difference between the N/Z of fragments in 
neutron rich and neutron poor sources, we see that it becomes slightly 
less for the high $E^{*}_s$. In particular, the difference of 
$(N/Z)_{52}-(N/Z)_{56}$ taken between the cases of $E^{*}_s$ = 6 and 10 AMeV 
and averaged for fragments with $A$ = 6-20, is 0.0051, i.e., the share of 
free neutrons increases a little bit with the excitation energy. 
\par
The secondary de-excitation of hot fragments changes their properties. 
Fig.~3 shows the N/Z ratios and relative yields obtained for the 
final cold fragments. Both masses and neutron content of fragments 
decrease considerably. The  decrease in the N/Z ratio is due to higher 
neutron evaporation probability compared to those for the protons.
The N/Z of light fragments vary with $Z$, 
reflecting their shell structure. However, as one can see from 
the middle panel, the qualitative difference between the N/Z ratios of hot 
fragments produced by sources with different isospin at the same excitation preserves after 
the secondary de-excitation, though it becomes smaller. 
Moreover, one can see that the solid and dotted lines in the middle 
panel are more distinguished than in the Fig.~2. The average difference 
between the cases with the two excitations increases up to 0.0168 for 
the fragments A = 6-20. This means that the difference of the N/Z of 
fragments produced in neutron rich and neutron poor sources decreases 
more for the high excitation energy, as a result of more intensive secondary de-excitation. 
\section{Comparison of the relative densities with SMM calculation}
We now compare the measured relative reduced neutron and proton 
densities for the Fe + Ni and Fe + Fe systems with the calculation. 
The properties of fragments discussed above are manifested in the evolution of these densities. 
Fig. 4 shows the relative densities plotted as a function of excitation 
energy of the source undergoing fragmentation. The excitation energies were 
obtained from the BNV calculations carried out for various beam energies. 
The error bars for the excitation energies 
corresponds to the two different equation of state used in the BNV calculation, namely, 
the asy-stiff and the asy-soft equation of state. The regions between solid 
lines shown in Fig. 4 are the SMM calculation of the densities obtained from the secondary fragment 
distribution. The regions between dotted lines 
correspond to the densities calculated from the primary fragment distribution. The width of the two regions 
represents the measure of sensitivity of the calculation to the assumed 
source size, i.e. for the sources with $A_s$ = 58 and 
$A_s$ = 116. The SMM calculations fit the experimental trends quite well, 
demonstrating a smooth decrease in $\rho_n/\rho^{Ni}_n$ with 
increasing excitation energy. 
    \begin{figure}
    \includegraphics[width=0.5\textwidth,height=0.50\textheight]{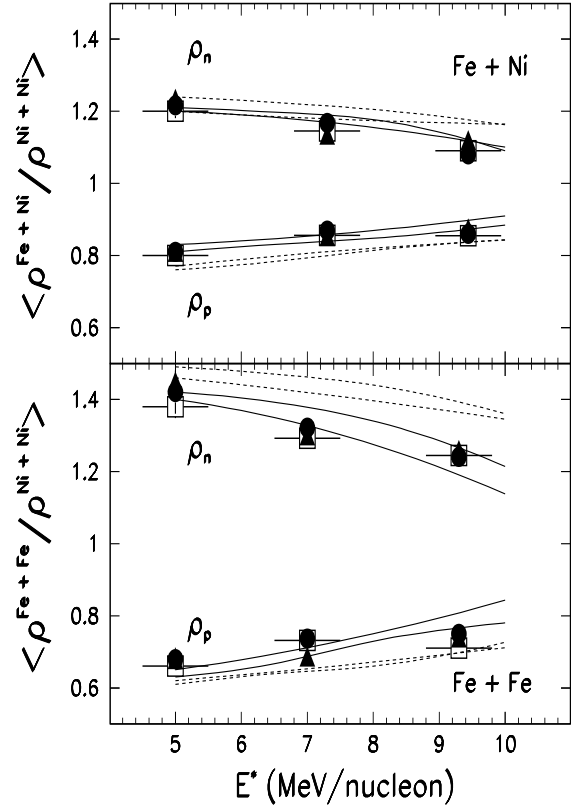}
    \caption{Relative reduced neutron and proton densities as a function of 
         excitation energy for Fe + Ni and Fe + Fe reactions. The errors bars for the 
         excitation energy is explained in the text. The different (solid and 
         dotted) regions correspond to the statistical model calculations as discussed in the text.} 
    \end{figure}
Both large- and small-size sources show very similar behavior indicating 
that the uncertainty of the source size does not influence the 
overall trends. 
\par
The calculation suggests the following interpretation of the data. 
The main difference in the relative densities is obviously caused by the 
difference of the N/Z of the sources. It is largest for the Fe + Fe case. 
The important observation is the decreasing of this difference with 
the excitation energy, demonstrated in \cite{SHE03}. 
In fact two effects contribute to it. The first one is a small decreasing 
of the difference between the N/Z ratios of primary fragments produced 
by the neutron-rich and neutron-poor sources with the excitation energy. 
This is because for high $E^{*}_s$ more neutron rich primary fragments are 
produced, however, if the small clusters are already neutron rich, 
free neutrons are predicted. 
Nevertheless, the secondary de-excitation is very important, since it changes 
drastically the N/Z of fragments. In the case of a larger internal 
excitation of fragments, the secondary decay produces more 
similar cold fragments, despite the difference in their initial N/Z ratios. 
This leads to decreasing of $\rho_n/\rho^{Ni}_n$ (and the 
corresponding increasing $\rho_p/\rho^{Ni}_p$), observed in 
the experiment. 
\section{Conclusion} 
We have explained the observed evolution of the fragment isospin 
with the excitation energy in the framework of the SMM. 
This is consistent with the following picture of 
the reaction: The primary fragments produced at the freeze-out density 
in the central collisions are 
excited and usually contain more neutrons than observed in cold 
fragments. Moreover, their N/Z ratio is only slightly 
less than the N/Z ratio of the excited nuclear systems. In our cases, 
the internal excitation energy of the primary fragments 
increases with the total excitation energy of the system. 
Therefore, the secondary de-excitation of fragments becomes 
more efficient in removing neutrons from fragments, and, the difference 
of the primary N/Z ratios disappears 
gradually with excitation energy, for the final cold fragments. 
However, these characteristics of primary fragments (their 
internal excitation energies and N/Z ratios), and the observed isospin 
evolution, are obtained for very high excitation energies, 
when the initial compound nucleus dissociates completely into small fragments. 
The isospin of fragments produced at energies near the multifragmentation 
threshold may behave differently \cite{BOT01}. 
We believe that the present results demonstrate that the properties 
of hot fragments can be effectively estimated by investigating general 
trends of cold fragments versus the excitation energy and the 
neutron richness of the sources. 
\section{Acknowledgment}
The authors wish to thank the staff of the Texas A$\&$M Cyclotron facility 
for the excellent beam quality. This work was supported in part by 
the Robert A. Welch Foundation through grant No. A-1266, and the Department 
of Energy through grant No. DE-FG03-93ER40773. 
One of the authors (ASB) thanks Cyclotron Institute TAMU for hospitality 
and support. 

\bibliography{isodens6.bbl}

\end{document}